%% file: WAA.tex
\documentclass[12pt,a4paper]{article}
\usepackage[utf8]{inputenc}
\usepackage{amsmath,graphicx,dcolumn,subcaption,mathtools,amssymb}
\usepackage{braket}
\usepackage{siunitx}
\usepackage{appendix}
\usepackage[mathscr]{eucal}
\usepackage{bm}
\usepackage{bbm}
\usepackage{cancel}
\usepackage{color}
\usepackage{rotating}
\usepackage{xcolor}
\usepackage[most]{tcolorbox}
\usepackage{multirow}

\usepackage[numbers,sort&compress]{natbib}
\usepackage[]{float}
\usepackage[font=small, labelfont=bf]{caption}

\usepackage{slantsc}
\usepackage{xspace}

\usepackage{setspace}
\allowdisplaybreaks

\usepackage{hyperref}
\usepackage{footnotebackref}

\input{definitions.tex}

\begin{document}
\sloppy
\hypersetup{pageanchor=false}
\begin{titlepage}
\begin{flushright}
  ZU-TH 80/25\\
  TUM-HEP 1579/25\\
\end{flushright}

\renewcommand{\thefootnote}{\fnsymbol{footnote}}
\vspace*{0.5cm}

\begin{center}
  {\Large \bf
    NNLO QCD predictions for $\Wgammagamma$ production at the LHC
  }
\end{center}

\par \vspace{2mm}
\begin{center}
  {\bf Paolo Garbarino}$^{(a)}$, {\bf Massimiliano Grazzini}$^{(a)}$,\\[0.2cm] {\bf Stefan Kallweit}$^{(a)}$ and {\bf Chiara Savoini}$^{(b)}$

\vspace{5mm}

$^{(a)}$ Physik Institut, Universit\"at Z\"urich, Winterthurerstrasse 190, 8057 Z\"urich, Switzerland\\[0.2cm]
$^{(b)}$ Physics Department, TUM School of Natural Sciences, Technical University of Munich, James-Franck-Stra{\ss}e 1, 85748 Garching, Germany

\vspace{5mm}

\end{center}

\par \vspace{2mm}
\begin{center} {\large \bf Abstract} 

\end{center}
\begin{quote}
  \pretolerance 10000

\input{WAA.abstract}

\end{quote}

\vspace*{\fill}
\begin{flushleft}
November 2025
\end{flushleft}
\end{titlepage}

\clearpage
\pagenumbering{arabic}
\setcounter{page}{1}
\hypersetup{pageanchor=true}


\renewcommand{\thefootnote}{\fnsymbol{footnote}}

\input{WAA.maintext}

\appendix
\newpage
\bibliography{biblio}

\end{document}

%% file: definitions.tex
\providecommand{\href}[2]{#2}

\newcommand\as{\alpha_{\mathrm{S}}}

\def\to{\rightarrow}

\newcommand\Matrix{{\sc Matrix}\xspace}

\newcommand\OpenLoops{{\sc OpenLoops}\xspace}
\newcommand\Recola{{\sc Recola}\xspace}

\newcommand\GeV{\ensuremath{{\rm GeV}}\xspace}
\newcommand\TeV{\ensuremath{{\rm TeV}}\xspace}

\newcommand\rcut{\ensuremath{r_{\mathrm{cut}}}\xspace}
\newcommand\qT{\ensuremath{q_T}\xspace}
\newcommand\muF{\ensuremath{\mu_F}\xspace}
\newcommand\muR{\ensuremath{\mu_R}\xspace}

\newcommand\Wgammagamma{\ensuremath{W\gamma\gamma}\xspace}

\newcommand{\pTgamma}{\ensuremath{p_{T,\gamma}}\xspace}

\newcommand{\absetagamma}{\ensuremath{|\eta_{\gamma}|}\xspace}

\newcommand{\pTmiss}{\ensuremath{p_{T,{\rm miss}}}\xspace}

\usepackage{etoolbox}
\makeatletter
\patchcmd{\@sect}{#8}{\boldmath #8}{}{}
\let\ori@chapter\@chapter
\def\@chapter[#1]#2{\ori@chapter[\boldmath#1]{\boldmath#2}}
\makeatother

%% file: WAA.abstract.tex
Triboson production processes play a crucial role in probing the electroweak sector
of the Standard Model, as they involve quartic gauge-boson couplings already at the
tree level. 
With these measurements entering the precision era at the Large Hadron Collider~(LHC),
accurate theoretical predictions become indispensable. We present the computation of
the next-to-next-to-leading-order (NNLO) QCD radiative corrections to
the production of a $W$ boson in association with two photons ($W\gamma\gamma$) at the LHC.
The calculation is exact, except for the finite part of the two-loop contribution,
which is included in the leading-colour approximation.
Predictions for the fiducial cross section and selected kinematic
distributions are provided at a centre-of-mass energy of \mbox{$\sqrt{s}=13$\,TeV},
under standard experimental selection cuts. 
In line with observations for other multiboson processes involving direct photons,
we find sizable NNLO corrections that enhance the
next-to-leading-order predictions by about $23\%$, with residual perturbative uncertainties
that can be roughly estimated to be at the $5\%$ level.

%% file: WAA.maintext.tex
\section{Introduction}
\label{sec:intro}
Over the past decade, precision measurements at the LHC have enabled detailed studies of
the trilinear couplings of electroweak~(EW) gauge bosons through diboson production.
To achieve a precise understanding of the EW symmetry breaking within the Standard Model~(SM),
the quartic couplings must likewise be scrutinised with a comparable level of accuracy.
This requires a more refined investigation of triboson production processes, which ---
together with diboson production via weak-boson scattering --- already probe quartic
gauge couplings at the tree level.
Deviations from SM predictions could thus directly point to effects of new physics
in the EW sector.
Moreover, triboson processes constitute a dominant background to other key SM processes.
For example, $V\gamma\gamma$ (\mbox{$V=Z,W$})
production is an irreducible background to $VH$ production via Higgsstrahlung,
with the Higgs boson decaying into two photons.
Despite their significantly smaller cross sections, triboson processes are now within reach
at the LHC, making reliable theoretical predictions essential.
Among the first processes of the triboson class, $V\gamma\gamma$ production
has been measured both at 8\,\TeV and 13\,\TeV by ATLAS and CMS, as reported in
Refs.~\cite{ATLAS:2015ify, ATLAS:2023avk} and~\cite{CMS:2017tzy, CMS:2021jji}, respectively.

On the theory side, next-to-leading-order~(NLO) QCD corrections to triboson production have
been computed mainly for leptonic decay
channels~\cite{Lazopoulos:2007ix,Hankele:2007sb,Binoth:2008kt,Campanario:2008yg,Bozzi:2009ig,Bozzi:2010sj,Bozzi:2011wwa,Baur:2010zf,Bozzi:2011en,Campbell:2012ft},
and have proven essential for reducing the gap between predictions and experimental
measurements.
Triboson production in SM Effective Field Theory~(SMEFT) at NLO QCD was studied in
Ref.~\cite{Celada:2024cxw}.
Complete NLO EW corrections, mostly including fully leptonic decays of the heavy
gauge bosons, are available for some relevant triboson
processes~\cite{Schonherr:2018jva,Dittmaier:2019twg,Cheng:2021gbx,Denner:2024ufg,Denner:2024ndl}.
Those involving two or more final-state photons, i.e.\ triphoton and
$V\gamma\gamma$ production, have been discussed in Ref.~\cite{Greiner:2017mft}.
NNLO QCD corrections are expected to be important, owing to the close similarity with
diboson production processes, where they have been shown to yield sizeable and positive
effects, especially in the presence of final-state
photons~\cite{Catani:2011qz,Grazzini:2013bna,Grazzini:2015nwa,Campbell:2016yrh,Grazzini:2017mhc,Catani:2018krb,Campbell:2017aul,Campbell:2021mlr}. 
Indeed, in the case of triphoton production --- the only triboson process for which NNLO QCD
predictions are currently available~\cite{Chawdhry:2019bji,Kallweit:2020gcp}%
\footnote{These predictions rely on a leading-colour approximation of the relevant
two-loop amplitudes. Complete amplitudes including subleading-colour contributions have
been made available in the meanwhile~\cite{Abreu:2023bdp}.}
--- very large NNLO corrections of $\mathcal{O}(60\%)$ relative to the NLO prediction
have been observed at the level of fiducial cross sections.
It was also shown that good agreement with the data measured by the ATLAS
collaboration~\cite{ATLAS:2017lpx} could only be achieved by including those corrections.

While tree-level and one-loop amplitudes can be obtained using automated tools,
a complete NNLO QCD computation is typically limited by the availability of the
corresponding two-loop amplitudes, whose complexity grows rapidly with the number
of (massive or off-shell) legs.
Until recently, in the class of triboson processes, only the two-loop amplitudes for
triphoton production, involving exclusively massless external legs, were available first
in the leading-colour approximation~(LCA)~\cite{Chawdhry:2019bji,Kallweit:2020gcp}
and later supplemented by all subleading-colour structures~\cite{Abreu:2023bdp}.
In Ref.~\cite{Badger:2024sqv}, the first computation of the two-loop amplitudes for a
triboson process with one off-shell leg, namely \Wgammagamma production, has been reported.
In this Letter, we present the first NNLO QCD predictions for \Wgammagamma,
exploiting these two-loop amplitudes and the implementation of the \qT-subtraction formalism~\cite{Catani:2007vq}
in the \Matrix framework~\cite{Grazzini:2017mhc}.%
\footnote{In this first NNLO computation
of \Wgammagamma production, we restrict ourselves to the LCA for the two-loop amplitudes.
The evaluation of the subleading-colour contributions, available upon request in
Ref.~\cite{Badger:2024sqv}, is extremely slow, and its inclusion is left for future work.}

\section{Computational details}
\label{sec:computation}

\begin{figure}
\centering
\includegraphics[width=0.48\textwidth]{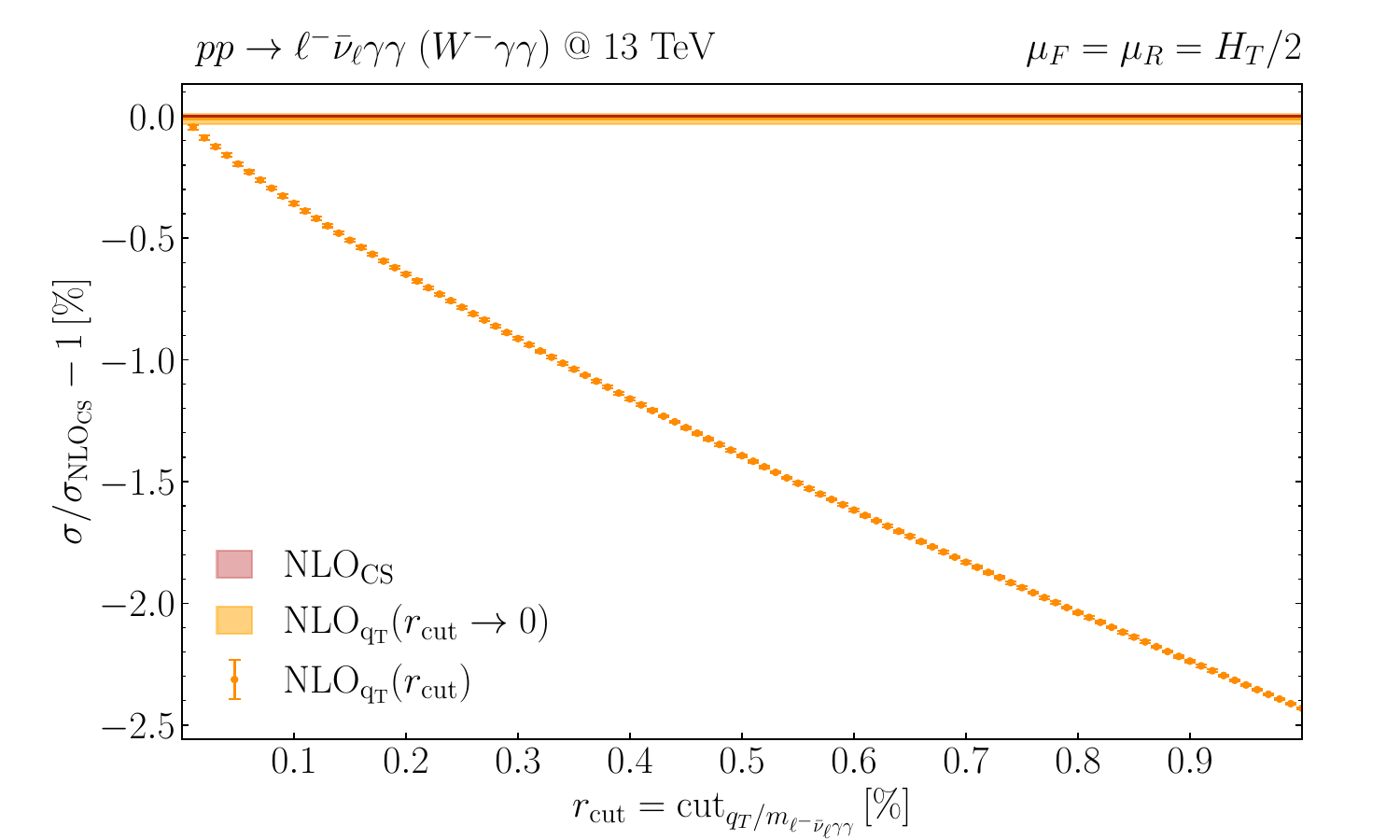}
\hfill
\includegraphics[width=0.48\textwidth]{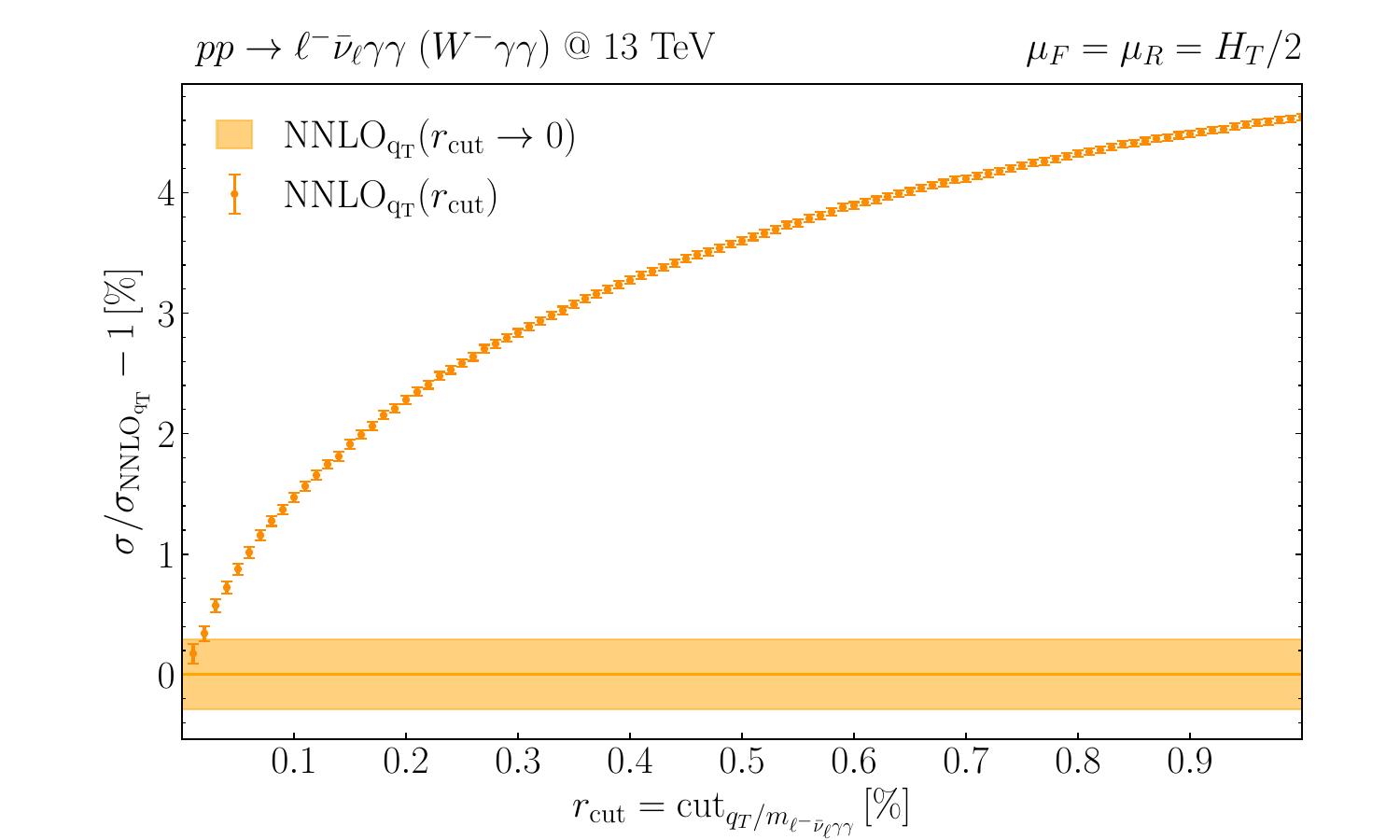}
\caption{\label{fig:qTcut} \rcut dependence of the NLO and NNLO cross sections for
\mbox{$pp\to \ell^-\bar{\nu}_\ell\gamma\gamma$} (dotted points with errorbars) and the
extrapolated result, \mbox{$\rcut\to 0$} (solid), at the central scale \mbox{$\muF=\muR$}.
}
\end{figure}

In this Section, we briefly discuss the procedure adopted for the NNLO QCD computation.
For brevity, we refer to the process as \Wgammagamma production; 
however, the calculation is performed for the complete leptonic final states,
without invoking any resonance approximation. 
To be precise, in this work we consider both the \mbox{$pp\to\ell^+\nu_\ell\gamma\gamma$}
and \mbox{$pp\to\ell^-\bar{\nu}_\ell\gamma\gamma$} processes.
Belonging to the class of colourless final-state systems, the treatment of infrared~(IR)
singularities can be addressed by applying the standard \qT-subtraction
formalism~\cite{Catani:2007vq}, as implemented in the \Matrix
framework~\cite{Grazzini:2017mhc}, which we have suitably extended to deal with massive
triboson production. 
All NLO-type singularities are treated using the dipole-subtraction
method~\cite{Catani:1996jh,Catani:1996vz}.
The NNLO computation in the \qT-subtraction formalism requires the introduction of a
technical cut-off, \rcut, on the dimensionless variable \mbox{$r=q_T/Q$}, where $q_T$ is
the transverse momentum of the colourless system and $Q$ its invariant mass.
The prediction for the NNLO fiducial cross section is then obtained by numerically
performing the extrapolation \mbox{$\rcut\to 0$}.
For distributions, this procedure is applied on a bin-wise level,
thereby providing differential results that are free from power corrections in \rcut.
As for the fiducial results, an extrapolation error is assigned.
Combined with the statistical error from the phase-space integration, this provides
the overall numerical uncertainty.
It is well known that the photon-isolation procedure induces a significant
\rcut dependence in the fiducial cross section, with the resulting power corrections
being larger for tighter isolation~\cite{Grazzini:2017mhc,Ebert:2019zkb}.
Nevertheless, as shown e.g.\ in Ref.~\cite{Kallweit:2020gcp} for triphoton production,
those power corrections are under control at the level of a few permille through
the extrapolation approach adopted in \Matrix.
In Figure \ref{fig:qTcut}, we explicitly display the NLO and NNLO cross sections for
$W^-\gamma\gamma$ production as a function of \rcut.
The extrapolated result for \mbox{$\rcut\to 0$} is also presented, together with the
corresponding extrapolation uncertainty (orange band), estimated via the standard
\Matrix procedure. For reference, at NLO we include the \rcut-independent
result (red line) obtained with dipole subtraction.
We find perfect agreement between the two subtraction methods at the sub-permille level
at NLO.
The numerical stability of our results allows us to conservatively assign a numerical
error of a few permille to the predicted fiducial NNLO cross section.

The computation within the \Matrix framework is fully automated. 
Also the evaluation of tree-level and one-loop amplitudes relies on automated tools
like \OpenLoops~\cite{Cascioli:2011va, Buccioni:2017yxi,Buccioni:2019sur}
and \Recola~\cite{Actis:2016mpe,Denner:2017wsf,Denner:2016kdg}. 
The only ingredient that needs to be treated on a process-by-process basis is represented
by the two-loop amplitudes. For \Wgammagamma, these amplitudes were recently made available
in the LCA~\cite{Badger:2024sqv}.
We have implemented the provided results into a dedicated \texttt{C++} library suitable
for numerical evaluation, building on and improving upon the amplitude libraries developed
for the calculations in Refs.~\cite{Buonocore:2022pqq,Buonocore:2023ljm,Devoto:2024nhl}.%
\footnote{The IR-finite remainder of the two-loop amplitude is computed according to the conventions of Ref.~\cite{Becher:2009cu}.}
To ensure the reliability and stability of the amplitudes despite the large numerical
cancellations expected at intermediate stages of the calculation,
we have introduced a sophisticated
numerical rescue system while keeping sustainable runtimes per evaluated phase-space point.
By default, we rely on quadruple-precision arithmetic throughout for the evaluation of the
rational coefficients of the master integrals, whereas the time-consuming evaluation of the
five-point one-mass master integrals is performed in double precision via the
\texttt{PentagonFunctions++}
library~\cite{git:PentagonFunctions,Chicherin:2020oor,Chicherin:2021dyp,Abreu:2023rco}.
The rescue system exploits the symmetry of the amplitude under the exchange of the two
photons. Thus, by evaluating the amplitude for both permutations
--- at the cost of doubling the runtime ---
we obtain a robust estimate of potential precision loss.
Beyond a certain user-defined threshold, the amplitude is re-evaluated at octuple precision
for the rational coefficients. Again, this re-evaluation can be performed for both photon
permutations, and if a reasonable agreement is still not reached, the precision for the
master-integral evaluation is increased to quadruple.
This procedure yields numerically stable amplitudes for all considered phase-space points,
with an average runtime of approximately six seconds, which is fully acceptable and does not
significantly impact the overall computational cost.

\section{Numerical results}
\label{sec:results}
In this Section, we present our predictions for the
\mbox{$pp\to \ell\nu_\ell\gamma\gamma$} processes at NNLO in QCD.
We consider proton--proton collisions at
a centre-of-mass energy of \mbox{$\sqrt{s}=13$\,TeV} and employ typical fiducial cuts
inspired by the recent ATLAS measurement
of \Wgammagamma production~\cite{ATLAS:2023avk}. 
We use a smooth-cone isolation \cite{Frixione:1998jh}, but slightly
adapt the standard threshold \mbox{$(E_{T,{\rm max}} = \epsilon p_{T,\gamma})$} to also account
for a constant term ($E_{T,\rm{thres}}$), i.e.\ we define
\mbox{$E_{T,{\rm max}}=\epsilon p_{T,\gamma}+E_{T,\rm{thres}}$}.
More precisely, for every \mbox{$r\leq R$},
where \mbox{$r=\sqrt{(\Delta\eta)^2+(\Delta\phi)^2}$} describes a cone of radius $r$ around
the photon ($\eta$ and $\phi$ denote pseudorapidity and azimuth, respectively),
we require \mbox{$E_{T,{\rm had}}(r)\leq E_{T,{\rm max}}\left(\frac{1-\cos r}{1-\cos R}\right)^n$},
with \mbox{$n=1$} and \mbox{$R=0.4$}, for the total amount of partonic transverse energy
inside the cone.
To resemble the fixed-cone criterion used in Ref.~\cite{ATLAS:2023avk},
we set \mbox{$\epsilon=0.032$} and \mbox{$E_{T,\rm{thres}}=6.53\,\GeV$}.

The isolated photons must
fulfill \mbox{$\pTgamma>20\,\GeV$} and \mbox{$\absetagamma<2.37$}.
The charged lepton is required to have \mbox{$p_{T,\ell}>25\,\GeV$} and
\mbox{$\left|\eta_\ell\right|<2.47$}, while the missing transverse momentum of the event
needs to fulfil \mbox{$\pTmiss>25\,\GeV$}.
The conditions \mbox{$\Delta R_{\gamma,\ell}>0.4$} and \mbox{$\Delta R_{\gamma,\gamma}>0.4$} are
imposed on the separation between the charged lepton and the photons and between the
two photons, respectively. We further impose the condition \mbox{$m_T^W>40\,\GeV$}
on the transverse mass of the $W$ boson, defined as
\mbox{$m_T^W=\sqrt{2p_{T,\ell}p_{T,\rm miss}(1-\cos\Delta\phi)}$} where $\Delta\phi$ is the
difference in azimuthal angles between the lepton momentum and
the missing transverse momentum $p_{T,\rm miss}$.

We employ the $G_\mu$ scheme, and set the input parameters to the
PDG values~\cite{Patrignani:2016xqp}: \mbox{$G_F = 1.16639\times 10^{-5}\,\GeV^{-2}$},
\mbox{$m_W=80.385\,\GeV$}, \mbox{$\Gamma_W=2.0854\,\GeV$},
\mbox{$m_Z = 91.1876\,\GeV$}, and \mbox{$\Gamma_Z=2.4952\,\GeV$}.
We apply the complex-mass scheme~\cite{Denner:2005fg},
and thus compute the EW mixing angle as
\mbox{$\cos\theta_W^2=\mu_W^2/\mu_Z^2$},
with \mbox{$\mu_V^2=m_V^2-i\Gamma_V\,m_V$} (\mbox{$V=W,Z$}),
as well as
\mbox{$\alpha=\sqrt{2}\,G_F \left|\mu_W^2\left(1-\mu_W^2/\mu_Z^2\right)\right|/\pi$}.
Following the reasoning of Ref.~\cite{Buccioni:2019sur}, we replace
one factor of $\alpha$ by \mbox{$\alpha_0=1/137.036$} for each identified final-state photon.
We use a diagonal Cabibbo--Kobayashi--Maskawa~(CKM) matrix.
We work in the 5-flavour scheme and employ the \texttt{NNPDF40}~\cite{NNPDF:2021njg} 
parton distribution functions~(PDFs), 
with densities and QCD coupling $\as$ ($\as(m_Z)=0.118$) evaluated at each corresponding order 
through the \texttt{LHAPDF} interface~\cite{Buckley:2014ana}.
We choose the central renormalisation and factorisation scales as
\mbox{$\muR=\muF=H_T/2=(E_{T,\ell\nu}+p_{T,\gamma_1}+p_{T,\gamma_2})/2$},
where \mbox{$E_{T,\ell\nu}=\sqrt{m_{\ell\nu}^2+p^2_{T,\ell\nu}}$}. 
The scale uncertainties are obtained through the customary procedure of independently
varying the renormalisation and factorisation scales by a factor of two around their
common central value, with the constraint \mbox{$0.5\leq \mu_R/\mu_F \leq 2$}.

Before presenting our numerical results, we recall some features of QCD radiative corrections
for multiboson processes involving direct photons.
Exceptionally large NLO and NNLO corrections are observed for
diphoton~\cite{Catani:2011qz,Campbell:2016yrh,Catani:2018krb},
triphoton~\cite{Chawdhry:2019bji,Kallweit:2020gcp} and
$W\gamma$~\cite{Grazzini:2015nwa,Campbell:2021mlr} production. 
All these processes are driven by quark--antiquark annihilation at the Born level.
The (anti)quark--gluon channel opening up at NLO leads to a significant correction,
since its large luminosity can compensate the ${\cal O}(\as)$ factor.
A related and complementary observation is that the LO cross section is often suppressed
in specific phase-space regions either for {\it kinematical} or {\it dynamical} reasons. 
In diphoton production with asymmetric cuts on the photon transverse momenta, for example,
part of the NLO enhancement can be traced back to the opening of the phase-space region
in which the softer photon is radiated close to its kinematic
cut~\cite{Catani:2011qz,Campbell:2016yrh,Catani:2018krb}. 
For triphoton production, the regions where the azimuthal separation of the leading and the
subleading/trailing photons is small open up only at NLO, due to the selection cuts on the
photon transverse momenta~\cite{Kallweit:2020gcp}.
In the case of $W\gamma$ production, the LO suppression has instead a {\it dynamical} origin:
the LO matrix element vanishes \cite{Mikaelian:1979nr} for a certain scattering angle of the
$W$ boson in the partonic centre-of-mass frame.
Such {\it radiation zero} is present also for $W\gamma\gamma$ production~\cite{Baur:1997bn},
if the photons become collinear,
leading to a strong suppression of the LO cross section, as discussed in the following.

Our numerical predictions for the fiducial cross sections are presented in
Table~\ref{tab:inclusive}, where we show the LO, NLO and NNLO results,
both combined and split into $W^+\gamma\gamma$ and $W^-\gamma\gamma$ production,
together with the respective numerical (in brackets) and scale-variation uncertainties.%
\footnote{Note that these numbers account for both the $e^\pm$ and $\mu^\pm$ decay channels.}
The impact of the NLO corrections is huge, enhancing the LO result by about a factor of 3.6.
The NNLO corrections further increase the NLO cross section by about 23\%,
exceeding the NLO scale-variation uncertainties by almost a factor of three.
As anticipated, the huge NLO $K$-factor can be explained by the opening of the
(anti)quark--gluon channel at this perturbative order, and by the LO suppression
due to the radiation zero.
Indeed, we find that the quark--antiquark channel%
\footnote{By quark--antiquark we refer to the \textit{diagonal} channel that contributes
already at the Born level, i.e.\ with a trivial CKM matrix, the $u\bar{d}$ and $d\bar{u}$
channels of the same isospin doublet for $W^+\gamma\gamma$ and $W^-\gamma\gamma$,
respectively.}
accounts for \mbox{$\sim 43\%$} of the NLO cross section, while the (anti)quark--gluon channel
provides the remaining \mbox{$\sim 57\%$}. 
This pattern is only mildly modified at NNLO, where the quark--antiquark and
(anti)quark--gluon channels amount to \mbox{$\sim39\%$} and \mbox{$\sim 55\%$} of the
NNLO cross section, respectively.
Among the new channels at NNLO, the gluon--gluon channel has a negative and
quite small (\mbox{$\sim -0.7\%$}) impact, while the off-diagonal (anti)quark-initiated
channels contribute the remaining \mbox{$\sim 7\%$} of the NNLO cross section.

\begin{table}[t]
\renewcommand{\arraystretch}{1.5}
\begin{center}
\begin{tabular}{clll}
${\sigma \rm [fb]}$ & \multicolumn{1}{c}{$W^+\gamma\gamma$} & \multicolumn{1}{c}{$W^-\gamma\gamma$} & \multicolumn{1}{c}{$W\gamma\gamma$} \\
\hline
$\rm LO$ & $2.011\phantom{()}^{+4.8\%}_{-5.7\%}$ & $1.596\phantom{()}^{+5.5\%}_{-6.5\%}$ & $\phantom{0}3.607\phantom{()}^{+5.1\%}_{-6.0\%}$ \\
$\rm NLO$ & $6.983\phantom{()}^{+7.8\%}_{-6.3\%}$ & $5.966\phantom{()}^{+8.2\%}_{-6.6\%}$ & $12.949\phantom{()}^{+8.0\%}_{-6.5\%}$ \\
$\rm NNLO$ & $8.55(2)^{+4.5\%}_{-4.0\%}$ & $7.33(2)^{+4.4\%}_{-4.0\%}$ & $15.88(3)^{+4.5\%}_{-4.0\%}$ \\
\end{tabular}  
\end{center}
\caption{\label{tab:inclusive} Fiducial cross sections for
\Wgammagamma production in the setup described in the main text.
LO, NLO and NNLO predictions for \mbox{$W^+\gamma\gamma, W^-\gamma\gamma$} and their
combination are stated with their seven-point scale variation uncertainties.
At NNLO, numerical uncertainties, including the systematic error from the
\mbox{$\rcut\to 0$} extrapolation,
are shown in brackets. All numbers account for both the $e^\pm$ and $\mu^\pm$ decay channels.}
\end{table}

We briefly comment on the perturbative uncertainties affecting the NNLO predictions reported
in Table~\ref{tab:inclusive}. 
NNLO is the first perturbative order at which all partonic channels contribute and the
corrections to the large (anti)quark--gluon channel, which opens up at NLO, are fully included.
Moreover, the radiation zero is already washed out at NLO (see below). 
Therefore, no further sizeable corrections are expected beyond NNLO. 
Although scale variations can only provide a lower bound on the true theoretical uncertainty,
we can roughly estimate the residual perturbative uncertainty to be at the $5\%$ level.
A direct comparison with the measured fiducial cross section reported by ATLAS,
\mbox{$\sigma=13.8\pm 1.1\text{\,(stat)}^{+2.1}_{-2.0}\text{\,(syst)}\pm 0.1\text{\,(lumi)\,fb}$}~\cite{ATLAS:2023avk},
shows good agreement with both our NLO and NNLO predictions, within $1\sigma$ of the quoted
experimental uncertainties.

\begin{figure}
\centering
\includegraphics[height=0.38\textheight]{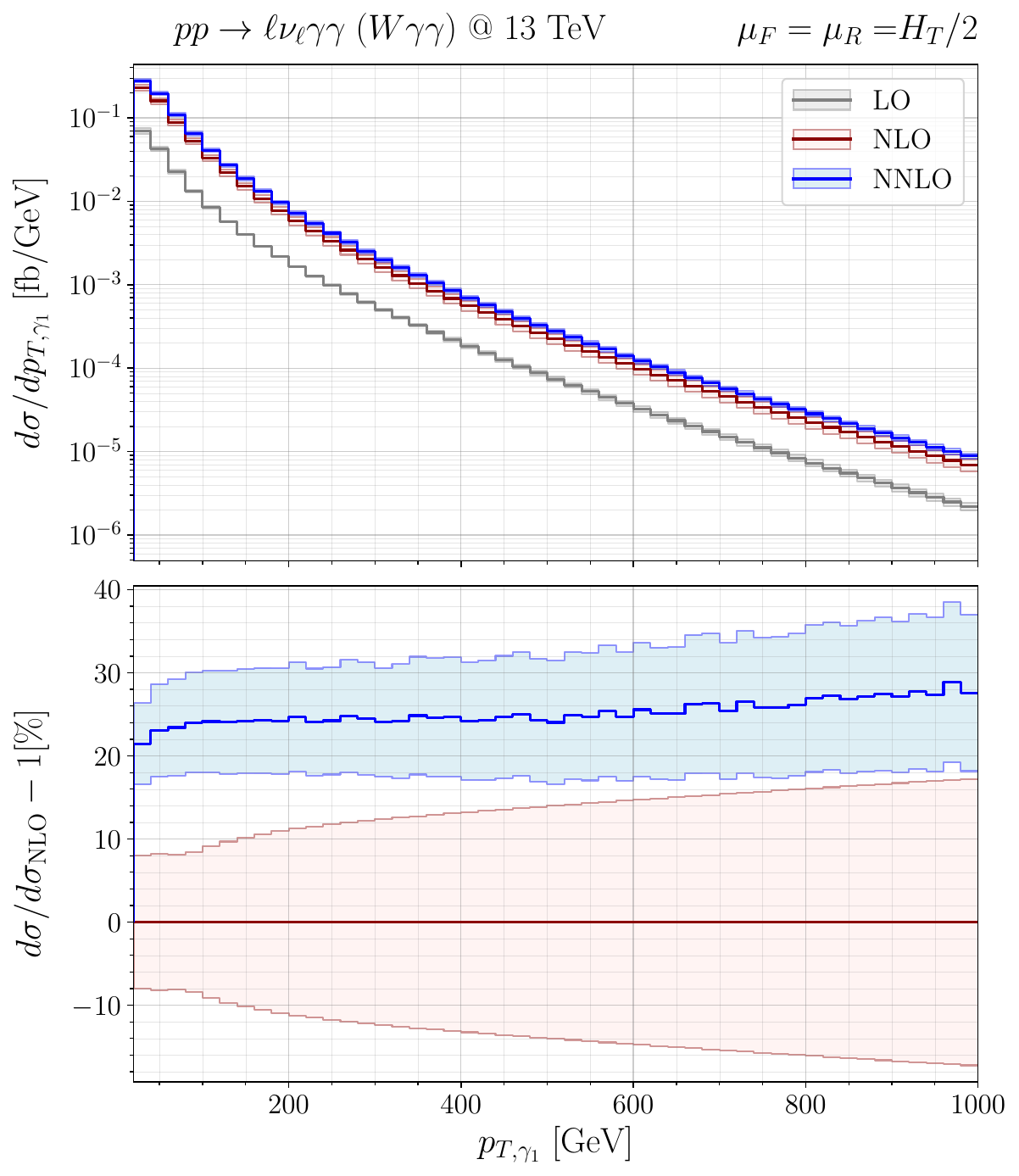}
\hfill
\includegraphics[height=0.38\textheight]{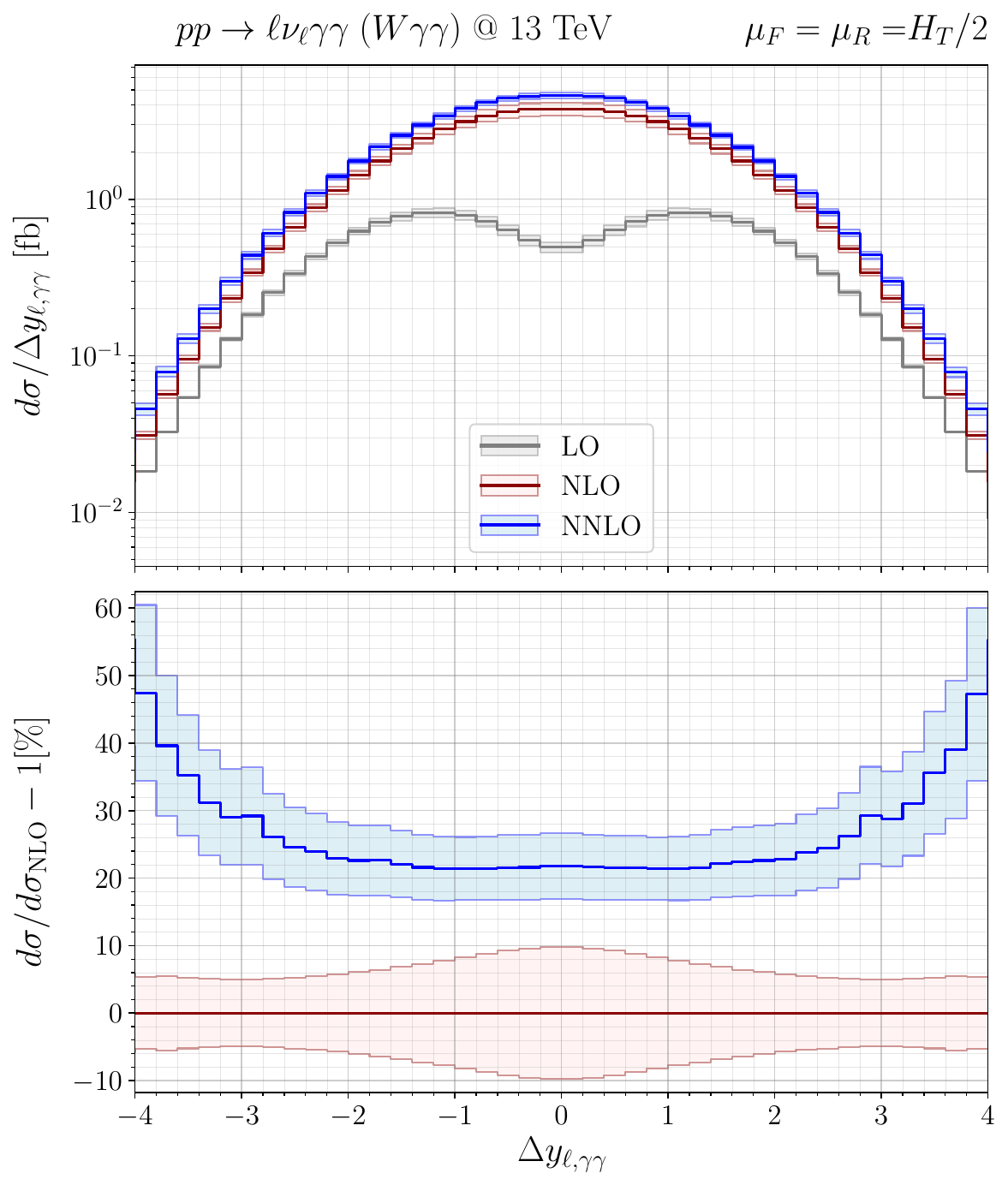}
\caption{\label{fig:diffresults} Distributions in the transverse momentum of the leading
photon, $p_{T,\gamma_1}$ (left), and in the difference in rapidities between the charged
lepton and the diphoton system, $\Delta y_{\ell,\gamma\gamma}$ (right), are presented for
\Wgammagamma production. The upper panels show the absolute prediction at LO, NLO and NNLO
accuracy, the lower panels relative corrections normalised by the NLO prediction, together
with their conventional (symmetrised) seven-point scale-variation bands.
}
\end{figure}

We now turn to the discussion of differential cross sections. 
In Figure \ref{fig:diffresults}, we
show the distributions for the transverse momentum of the leading photon, $p_{T,\gamma_1}$
(left plot), and the rapidity difference between the charged lepton and the diphoton pair,
$\Delta y_{\ell,\gamma\gamma}$ (right plot).
For each distribution, the upper panel displays the absolute predictions, while the lower panel
shows the ratios relative to the NLO result. The bands, obtained through the customary
seven-point scale variations, are symmetrised. To be precise, we take the maximum among the
upward and downward variations and assign it symmetrically to construct our final uncertainty,
leaving the central prediction unchanged.
The $p_{T,\gamma_1}$ distribution receives almost flat NNLO corrections of \mbox{$\sim20-30\%$} on
top of the NLO result. The NNLO (NLO) scale uncertainties range
from \mbox{$\sim4\%$} ($8\%$) at low $p_{T,\gamma_1}$ to \mbox{$\sim9\%$} ($17\%$) in the
high-$p_{T,\gamma_1}$ region.
In the low-$p_{T,\gamma_1}$ region, mostly contributing to the bulk of the cross section,
the NLO and NNLO bands do not overlap, consistent with our findings for the fiducial
cross sections. 
On the contrary, the bands become closer in the high-$p_{T,\gamma_1}$ tail and almost overlap
at \mbox{$p_{T,\gamma_1}\sim 1$\,TeV}.

The $\Delta y_{\ell,\gamma\gamma}$ distribution exhibits the well-known radiation-zero
pattern~\cite{Baur:1997bn},
responsible for a dip in the LO distribution around \mbox{$\Delta y_{\ell,\gamma\gamma} = 0$}.
This dip is completely washed out by real-emission contributions beyond LO.
Indeed, it is already filled at NLO,
with a correction amounting to more than $600\%$ and reflected by the widening of the
NLO scale-variation band in the central region.
In line with the fiducial NNLO $K$-factor, we find that the NLO and NNLO bands do not
overlap in the full range.

We finally discuss the impact of the two-loop virtual corrections on the NNLO
result. At the fiducial level, they amount to about $1\%$ of the full NNLO prediction.
At the differential level, their effect on the $p_{T,\gamma_1}$ distribution is about $1.5\%$ at the peak, decreasing
rapidly to the permille level in the tail. In contrast, for the $\Delta y_{\ell,\gamma\gamma}$ distribution, their impact amounts 
to a few permille in the bulk region, but increases up to $7\%$ in the strongly suppressed large-$\Delta y_{\ell,\gamma\gamma}$ region.
Overall, assuming that the relative accuracy of the LCA on the two-loop virtual
contribution is ${\cal O}(20\%)$, 
we can conclude that the ensuing uncertainty on the NNLO predictions is below $1\%$,
much smaller than the residual perturbative uncertainty.

\section{Summary}
\label{sec:summary}
In this Letter, we have presented the first NNLO QCD predictions for the production of
a $W$ boson in association with two photons (\Wgammagamma) at the LHC. The calculation has
been carried out within the process-independent implementation of the
$\qT$-subtraction method in the \Matrix framework.
Our results are exact, except for the finite part of the two-loop
amplitudes, recently computed in the
leading-colour approximation, which we have implemented in a dedicated
\texttt{C++} library for efficient numerical evaluation.
We find that the well-known radiation zero leads to significant NLO corrections
at the fiducial level, while NNLO corrections increase the NLO prediction by about 23\%
and reduce the perturbative uncertainty to the 5\% level.
We have studied the impact of QCD radiative corrections for the transverse-momentum
distribution of the leading photon and
the rapidity separation between the charged lepton and the diphoton system. The latter
provides direct sensitivity to the radiation-zero pattern, which is clearly
visible at LO, but completely washed out once higher-order corrections are included.
Our results provide the first predictions for a triboson process with
one off-shell leg at NNLO QCD accuracy. They will enable more stringent tests
of the SM, especially in future measurements of quartic gauge couplings as experimental
precision improves.

\section*{Acknowledgments}
We would like to thank Heather Russell for useful correspondence on Ref.~\cite{ATLAS:2023avk},
Vasily Sotnikov for advice on and immediate improvements in the
\texttt{PentagonFunctions++} library. We are further indebted to Simone Zoia for
discussions about the two-loop amplitudes exploited in this project.
This work is supported in part by the Swiss National Science Foundation (SNF) under
contract $200020\_219367$.
The work of C.S. is partly supported by the Excellence Cluster ORIGINS, funded by the
Deutsche Forschungsgemeinschaft (DFG, German Research Foundation) under
Germany's Excellence Strategy --- EXC-2094-390783311.